\documentstyle[prl,aps,floats,epsf]{revtex}

\begin{document}
\draft

\twocolumn[\hsize\textwidth\columnwidth\hsize\csname
@twocolumnfalse\endcsname
\title{Abrupt appearance of the domain pattern and fatigue of thin
ferroelectric films}
\author{A.M. Bratkovsky$^{1}$ and A.P. Levanyuk$^{2}$}
\address{$^{1}$Hewlett-Packard Laboratories, 3500 Deer Creek Road, Palo Alto,\\
California 94304\\
$^{2}$Departamento de F\'{i}sica de la Materia Condensada, CIII, Universidad\\
Aut\'{o}noma de Madrid, 28049 Madrid, Spain}
\date{August 11, 1999}
\maketitle

\begin{abstract}

We study the domain structure in ferroelectric thin films with a
`passive' layer (material with damaged ferroelectric properties) at
the  interface between the film and electrodes within a continuous
medium approximation. An abrupt transition from a monodomain to a
polydomain state has been found with the increase of the `passive'
layer thickness $d$. The domain width changes very quickly at the
transition (exponentially with $d^{-2}$). We have estimated  the
dielectric response $dP/dE$ (the slope of the hysteresis loop)  
in the `fatigued' multidomain state and found that it is in agreement
with experiment, assuming realistic parameters of the layer. 
We derive a simple universal relation for the dielectric response,
which scales as $1/d$, involving only the properties of the passive
layer. This relation qualitatively reproduces the evolution of the
hysteresis loop in fatigued samples and it could be tested with
controlled experiments.  
It is expected that the coercive field should increase with decreasing lateral
size of the film. 
We believe that specific properties of the domain structure under bias
voltage in ferroelectrics with a passive layer can resolve the
long-standing `paradox of the coercive field'.


\end{abstract}
\vskip 2pc ] 

\narrowtext
Recent studies of thin ferroelectric films have revived interest in
properties of the ferroelectric domain structures. It became clear that some
basic aspects of these properties remain unexplored, whereas they are of key
importance for applications. For instance, the progressive loss of
switchable polarization after repeated switching cycles, i.e. polarization
fatigue, is a serious problem in device applications of ferroelectrics. In
spite of extensive studies, the physics of the fatigue remains poorly
understood. Various mechanisms were considered over the years\cite
{land59,dpin,mil90,taylor95,emig1,lem96,tag96,gruv96}. In many cases the
deterioration of the switching behavior, like the loss of the coercive force
and of the squareness of hysteresis loop, were attributed to the growth of a
`passive layer' at the ferroelectric-electrode interface \cite
{land59,mil90,lem96}, or to the pinning of domain walls by defects\cite
{dpin,taylor95}. In this paper, we study the effects of a passive layer
(material with damaged ferroelectric properties). We assume an ideal
infinite ferroelectric film and treat it within the continuous medium
approximation. The model exhibits very interesting properties relevant for
real systems.

The passive layer leads to the appearance of a depolarizing field in the
ferroelectric, so that the system would tend to transform to a polydomain
state in order to reduce the field. There is, therefore, a transition
between the monodomain state, when the thickness $d$ of the passive layer is
zero, to a polydomain state otherwise. What is the nature of this
transition? This is the main theoretical question we address in this paper.

For a short-circuited infinite ferroelectric plate, as we shall show, the
domain structure exists for {\em any} value of the passive layer thickness,
in disagreement with some earlier results \cite{chensky82}. A surprising
feature of the present study is that the formation of the polydomain state
sets in {\em abruptly} with the appearance of the passive layer. The period
of the domain structure tends to infinity {\em exponentially} when the
passive layer thickness $d$ goes to zero. The theory allows one to study the
evolution of the domain structure in the asymptotic regime of very wide
domains and calculate its dielectric response.

The slope of the resulting $P-E$ (polarization - electric field) hysteresis
curve at $E=0$ is directly related to the thickness $d$ and dielectric
constant $\epsilon_g$ of the non-ferroelectric layer. We show that there is
a universal relation (\ref{eq:dPdV}) $dP/dE\propto \epsilon_g/d$.
This relation is in good correspondence with available experimental data. It
is important that even {\em without} pinning by defects, the response of
this structure to an external bias voltage is {\em rigid}. The implication
for real systems is that with the growth of the passive layer the hysteresis
loop very quickly deteriorates and {\em looses its squareness.}

The properties of the domain structure in the presence of a passive layer
are important for the problem of {\em switching} in ferroelectric thin
films. If we were to apply the external field in the {\em opposite}
direction to the spontaneous polarization in a monodomain state, the
equilibrium size of the domain with the `wrong' polarization would
eventually become {\em less} than the film size. Then a new domain wall
would appear in the film and we get a {\em switching}. Obviously, the
coercive field for this mechanism should then increase with decreasing {\em %
lateral } size of the film.

We shall consider a ferroelectric film under a bias voltage $U$ with
thickness $l$ separated from the top and bottom electrodes by passive layers
with thickness $d/2$ (Fig.~1, inset). 
For a system with a given {\em voltage}
on the electrodes, the equilibrium distribution of charges and polarization
corresponds to the extremum of\ the thermodynamic potential \ $\tilde{F},$\
which includes explicitly the work produced by the voltage source(s) and the
energy of the electric field\cite{LL8,BLlong,chensky82}: 
\begin{equation}
\tilde{F}=F_{{\rm LGD}}+\int dVE^{2}/(8\pi )-\sum_a^{\rm
electrodes}e_{a}\varphi _{a} 
\label{eq:LGD}
\end{equation}
Here $F_{{\rm LGD}}$ is the standard Landau-Ginzburg-Devonshire (LGD)
functional at {\em zero} electric field \cite{LL8,strukov}, $\vec E$ is the
electric field, while $e_{a}$ ($\varphi _{a})$ is the charge (potential) on
the electrode $a$. The last term in Eq.~(\ref{eq:LGD}) accounts for the work
of the external voltage source(s).

We are interested in a case when the ferroelectric has a spontaneous
polarization $\vec{P}_s\parallel z$ (Fig.~1, inset). The film has a
dielectric constant $\epsilon _{c}$ ($\epsilon _{a})$ in the $z$-direction
(in the $xy$-plane), and the dielectric constant of the passive layer is $%
\epsilon _{g}.$\ We select the $x$-axis perpendicular to the domain walls.
The potential $\varphi $ ($\vec{E}=-\nabla \varphi $) satisfies the
following equations in the ferroelectric and the passive layer, 
\begin{equation}
\epsilon _{a}\frac{\partial ^{2}\varphi _{f}}{\partial x^{2}}+\epsilon _{c}%
\frac{\partial ^{2}\varphi _{f}}{\partial z^{2}}=0,\quad \frac{\partial
^{2}\varphi _{g}}{\partial x^{2}}+\frac{\partial ^{2}\varphi _{g}}{\partial
z^{2}}=0,  \label{eq:figap}
\end{equation}
with the boundary conditions $\varphi =-(+)U/2,\quad z=+(-)(l+d)/2$, and 
\begin{equation}
\epsilon _{c}\frac{\partial \varphi _{f}}{\partial z}-\epsilon _{g}
\frac{\partial \varphi _{g}}{\partial z}=4\pi \sigma (x),\quad \varphi
_{f}=\varphi _{g}, ~ {\rm at}~ z=l/2,
\end{equation}
where the subscript $f$ ($g)$ denotes the
ferroelectric (passive layer, or a 
vacuum gap). Here $\sigma $ is the density of the bound charge due to
spontaneous polarization at the ferroelectric-passive layer interface, $%
\sigma =P_{ns}=\pm P_s$, depending on the normal direction of the
polarization at the interface, alternating from domain to domain,
Fig.~1(inset). We have assumed a usual separation of linear and
spontaneous polarization,  so that the displacement vector is $D_{i}=\epsilon
_{ik}E_{k}+P_{si}$, where $i,k=x,y,z$, and the dielectric response $\epsilon
_{ik}$ in uniaxial, Fig.~1(inset).
\begin{figure}[t]
\epsfxsize=3.8in
\epsffile{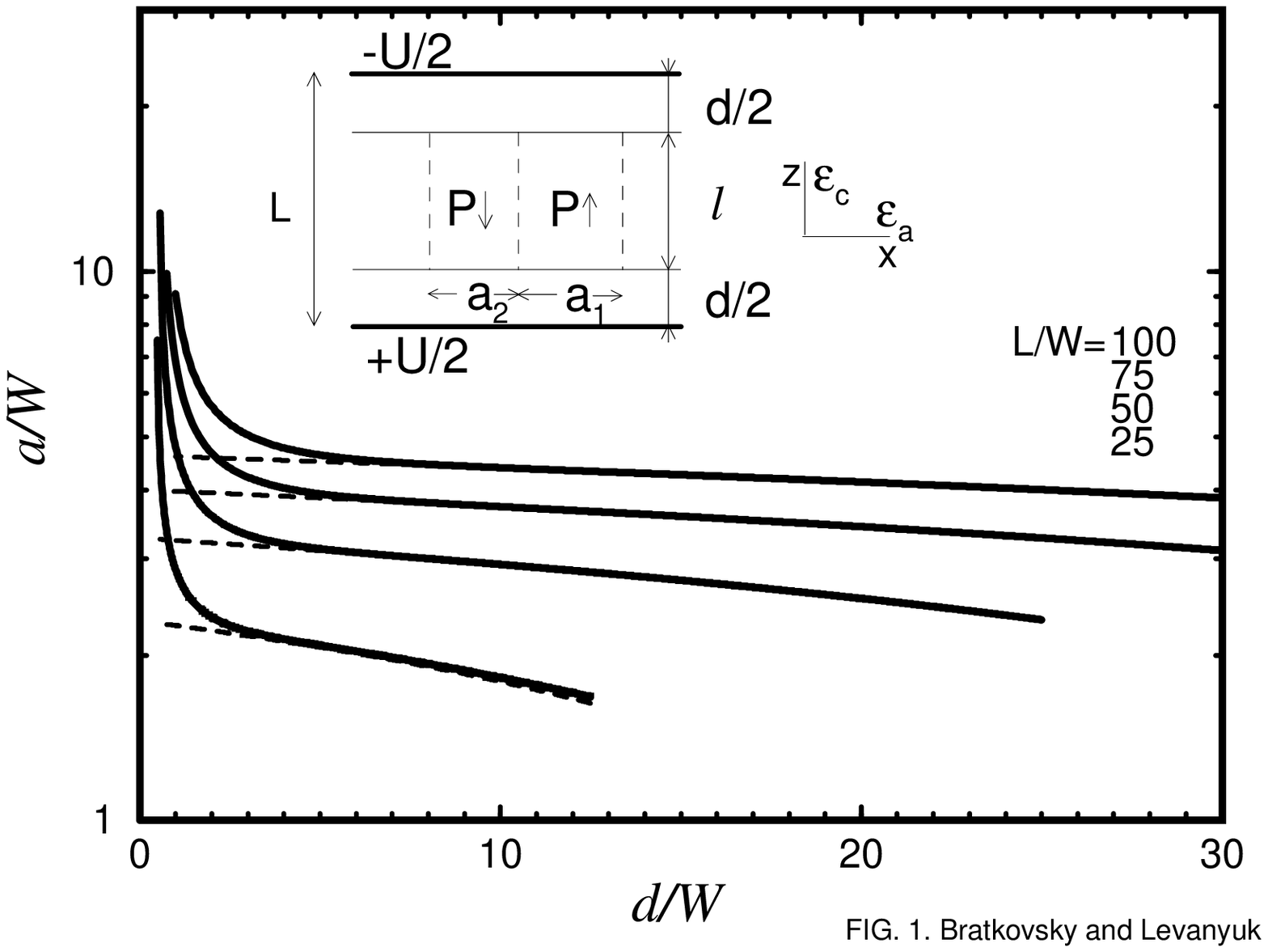}
\vspace{.02in}
\caption{ The domain width $a$ in a ferroelectric capacitor versus the
passive layer thickness $d$ for different separations between the electrode
plates $L$, $\protect\epsilon_g=\protect\epsilon_a=10$, $\protect\epsilon%
_c=200$ at zero bias. Dashed lines show the behavior of the Kittel width $%
a_K $. All quantities are scaled to $W$, the domain wall thickness [12].
Inset shows schematics of the electroded ferroelectric film (capacitor)
under bias voltage $U$. Note that the transition to very wide domains at
small $d$ is very sharp, even when shown on a logarithmic scale. }
\label{fig:1}
\end{figure}

We shall assume that the film is split into domains with alternating $z$%
-component of polarization and widths $a_{1}$ and $a_{2}$ so that the
pattern is periodic with a period $T=a_{1}+a_{2}$. This assumption is
perfectly reasonable, since the bound charges at the boundary between the
ferroelectric and the passive layer are fully compensated by the image
charges at the electrodes. In this situation the domains would not branch
(see e.g. \cite{strukov}, p.219). The solution of the Poisson equations (\ref
{eq:figap}) is then readily found by Fourier transformation 
\begin{eqnarray}
\sigma (x) &=&\sum_{n=-\infty }^{\infty }\sigma _{n}e^{\frac{2\pi inx}{T}},
\\
\sigma _{n} &=&\frac{P_s}{i\pi n}\left[ 1-\exp \left( \frac{2\pi ina_{1}}{T}%
\right) \right] ,\ \hspace{0.2in}n\neq 0, \\
\sigma _{n=0} &\equiv &\sigma _{0}=P_s\frac{a_{1}-a_{2}}{a_{1}+a_{2}}\equiv
P_s\delta ,  \label{eq:sig0} \\
\varphi _{\alpha }(x,z) &=&\sum_{n=-\infty }^{\infty }\varphi _{n\alpha
}(z)e^{\frac{2\pi \imath nx}{T}},  \label{eq:fifour}
\end{eqnarray}
where $n$ is the number of the Fourier component, and $\alpha =f,$ $g$. With
the assumption that $P_s={\rm const}$ inside the domains, 
the free energy
Eq.~(\ref{eq:LGD}) becomes $\tilde{F} = F_0+\frac{1}{2}\int_{\text{%
FE}}d{\cal A}\sigma \varphi $ $ - \frac{1}{2}\sum_{a}
e_{a}\varphi _{a}$, where $F_0=F_{\rm LGD}[P_s]$ is the free energy of the
ferroelectric with the polarization $P_s$ which 
contains the surface energy of the domain walls, and the integral is
taken over the interface between the 
ferroelectric and the passive layer. It is useful to
treat the homogeneous part ($n=0$) of the potential and the surface
charge separately from other terms. Then the result for the
(dimensionless) free energy per unit volume is 
\begin{eqnarray}
f &\equiv &\frac{\tilde{F}}{{\cal A}LP_s^{2}}=\frac{\delta }{2}\frac{\left(
4\pi \delta d/L\right) -\epsilon _{g}u}{\left( \epsilon _{c}d/l\right)
+\epsilon _{g}}  \nonumber \\
&&-\frac{ \epsilon_g  u }{ 8\pi }\frac{\left( \epsilon _{c}uL/l\right)
+4\pi \delta }{\left( \epsilon _{c}d/l\right) +\epsilon _{g}}%
+f_{K}(t)+f_{x}(t,\delta ,u),  \label{eq:ftild}
\end{eqnarray}
where the dimensionless bias voltage is $u\equiv U/P_sL$, and the first two
`homogeneous' terms describe the growth of the domains with the `right'
polarization (along the external field) at the expense of domains with
opposite polarization, $f_{K}(t)$ is the Kittel-like free energy \cite
{kit46,LL44} defining the dimensionless period of the domain structure $%
t\equiv (a_{1}+a_{2})/L$, and $f_{x}(t,\delta ,u)$ is the cross-term \cite
{BLlong} describing, in addition to the first two terms, the
appearance of a net polarization in the film under 
bias, while $L=l+d$ is the distance between the electrodes, Fig.~1(inset).
To find the equilibrium distribution of charges and polarization, this
functional should be varied with respect to $\delta$ and $t$ at {\em
constant} bias voltage (parameter $u$).

The domain structure with the passive layer can be determined
simply from a minimum of $f_{K}$ in the case of zero bias voltage, when $%
a_{1}=a_{2}=a,$ and 
\begin{eqnarray}
&&Lf_{K}=\frac{\Delta l}{a}+\frac{32a}{\pi ^{2}}\sum_{j=0}^{\infty }\frac{1}{%
(2j+1)^{3}}  \label{eq:fK} \\
&&\times \frac{1}{\epsilon _{g}\coth \left[ \frac{(2j+1)\pi d}{2a}\right] +%
\sqrt{\epsilon _{a}\epsilon _{c}}\coth \left[ \left( \frac{\epsilon _{a}}{%
\epsilon _{c}}\right) ^{1/2}\frac{(2j+1)\pi l}{2a}\right] }.  \nonumber
\end{eqnarray}
For a free standing film, $d\gg l$, Eq.~(\ref{eq:fK}) has a standard
(generalized Kittel) solution 
\begin{eqnarray}
a &=&a_{K}\equiv \left[ \pi ^{2}\left( \epsilon _{g}+\sqrt{\epsilon
_{a}\epsilon _{c}}\right) \Delta l/28\zeta (3)\right] ^{1/2}  \nonumber \\
&=&\left[ 0.3\left( \epsilon _{g}+\sqrt{\epsilon _{a}\epsilon _{c}}\right)
\Delta l\right] ^{1/2},  \label{eq:aK}
\end{eqnarray}
where $\Delta $ is the (temperature dependent) characteristic length,
which is proportional to the 
product of the inverse of the dielectric susceptibility $\chi $\ in the FE
phase and the domain wall thickness\ $W$ \cite{LL44,strukov}.

Most interesting is the ferroelectric film with a narrow passive layer, $%
d\ll l.$ \ When $\left( \frac{\epsilon _{a}}{\epsilon _{c}}\right) ^{1/2}%
\frac{\pi l}{2a}\gtrsim 1$, the second $\coth $ in (\ref{eq:fK}) can be
replaced by unity and neglected altogether in all terms in the sum with $%
j\lesssim a/(\pi d)$, since the first factor in the denominator is $\epsilon
_{g}\coth \left[ \frac{\pi d}{2a}(2j+1)\right] \sim \frac{2a}{\pi d}\gg 1.$
We see that the factor $a$ {\em cancels out} from all those terms in the
sum. The dependence of the second term (the so-called energy of emergence
per unit area) on $a$ becomes very weak, and the free energy minimum would
correspond to {\em very large domain widths}, $a\gg a_{K}$, Fig.~1. The
domain size may exceed the lateral size of the sample, and then we would
have a monodomain film. This surprising behavior is the result of the
effective {\em screening} of the bound charges on the ferroelectric-passive
layer interface by the metallic electrodes\cite{MagDomains}.

To analyze the transformation between monodomain and multidomain states, we
first consider an exactly solvable model with $\epsilon _{g}=\sqrt{\epsilon
_{a}\epsilon _{c}}$ in the regime $l/a\gtrsim 1.$ Then 

\begin{eqnarray}
Lf_{K} &=&\Delta l/a+\frac{32a}{\pi ^{2}\epsilon _{g}}\sum_{j=0}^{\infty }%
\frac{1}{(2j+1)^{3}}\frac{1}{\coth \left[ \frac{(2j+1)\pi d}{2a}\right] +1} 
\nonumber \\
&=&\Delta l/a+\frac{16d}{\pi \epsilon _{g}}I,\quad \text{where}
\nonumber \\
I &\equiv &\int_{0}^{1}{\rm d}\lambda \sum_{j=0}^{\infty }\frac{e^{-\lambda
(2j+1)\pi d/a}}{(2j+1)^{2}}=\frac{\pi ^{2}}{8}-\frac{\xi }{4}\ln \frac{e^{p}%
}{\xi },
\end{eqnarray}
up to terms $\sim \xi^3$,
where $\xi \equiv \pi d/a\ll 1,$ and $p=(3+\ln 4)/2$\cite{asymp}. Minimizing
the free energy $f_{K}$, we finally obtain the width of the domains in the
case of a narrow passive layer
\begin{equation}
a=\frac{\pi d}{2e^{1/2}}\exp \left( \frac{\epsilon _{g}\Delta l}{4d^{2}}%
\right) =0.95d\exp \left( 0.4\frac{a_{K}^{2}}{d^{2}}\right)   \label{eq:ad}
\end{equation}
This unusual solution is extremely sensitive to the thickness $d$ of the
passive layer, so that the domain width increases explosively for narrowing
passive layer. This conclusion is general, it applies at $a\gtrsim l$ too,
since the cancelation of $a$ persists in this case as well, as illustrated
by the results of the exact calculation in Fig. 1. Note that a similar case
of a ferroelectric capacitor with non-ferroelectric insulating layer at the
electrode surface has been considered recently \cite{watan98}, but the
exponential growth of domain thickness (\ref{eq:ad}) was overlooked, and
only the standard Kittel expression was obtained.

We have seen that the ferroelectric film splits into domains as a result of
the growth of a passive layer at the electrode surfaces. This substantially
decreases the coercive field. It will be defined by only the strength of \
the domain wall pinning. We can assess the validity of this scenario of
fatigued state by considering an ideal case with no pinning and comparing
the estimated slope of the hysteresis loop with experiment. To this end, we
shall estimate the value of $dP_{a}/dE$, where $P_{a}$ is the apparent
(measured) polarization (surface charge) $P_{a}=\sigma _{0}=P_s\delta $,
Eq.~(\ref{eq:sig0}), and $E$ is the electric field (in fact, it is more
convenient to find $dP_{a}/dU=L^{-1}dP_{a}/dE$). The linear response to bias
voltage is obtained from Eq.~(\ref{eq:ftild}), which, for small bias $U$,
takes the form 
\begin{equation}
f=f_{0}(t)+\frac{1}{2}S\delta ^{2}-Ru\delta ,  \label{eq:stiff}
\end{equation}
where $S$ is the domain structure stiffness with respect to changes in
apparent polarization $P_{a}$, and the coefficient $R$ defines the response
to the bias voltage. The slope of the hysteresis loop $L(dP_{a}/dU)=R/S$ is
shown in Fig. 2. Since the cross-term $f_{x}$ in Eq.~(\ref{eq:ftild})
changes only the numbers, we can obtain a simple analytical
approximation by neglecting this term to the first approximation. 
\begin{figure}[t]
\epsfxsize=3.8in
\epsffile{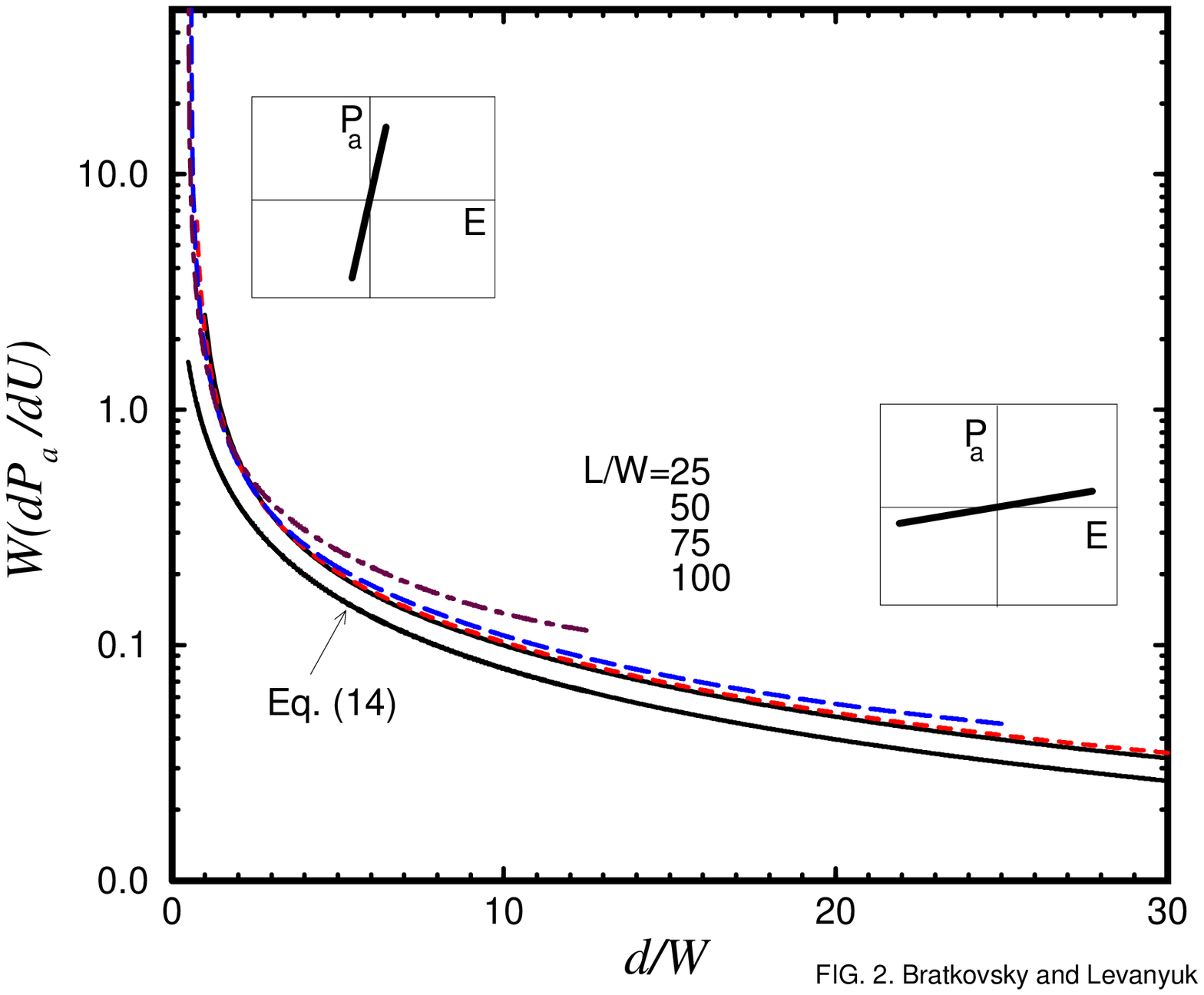}
\vspace{.02in}
\caption{ The calculated slope of the hysteresis loop, $W(dP_a/dU)$, in a
fatigued sample in comparison with the approximate expression, Eq.~(\ref
{eq:dPdV}). The parameters of the film are the same as in Fig.~\ref{fig:1}.
Insets show schematics of the very fast decrease of the slope of the
hysteresis $P_a-E$ loop with the growth of the passive layer (fatigue). }
\label{fig:2}
\end{figure}
The result is 
\begin{equation}
\frac{dP_{a}}{dU}=\frac{\epsilon _{g}}{4\pi d}.  \label{eq:dPdV}
\end{equation}
This remarkably simple relation is universal, and depends only on properties
of the passive layer [the expression reads $\epsilon _{0}\epsilon_g/d$
in SI units]. It is also rather accurate in comparison with 
the exact numerical calculation, Fig.~2. It is well known from experiment
that, as fatigue progresses ($d$ grows), the slope of the hysteresis loop
steadily decreases\cite{mil90,gruv96,tag95}. This behavior follows
immediately from our Eq.~(\ref{eq:dPdV}), the slope of hysteresis loop
decreases very rapidly with increasing $d$, Fig.~2. From experiment \cite
{tag95}, we can determine the slope in PZT\ films as $dP_{a}/dU\sim 15{\rm %
\mu Ccm}^{-2}{\rm V}^{-1}.$ One can estimate the corresponding thickness of
the passive layer from Fig. 2. Equation~(\ref{eq:dPdV})\ gives for the
passive layer thickness $d\sim 2-7$~nm with $\epsilon _{g}=3-10$, the
thickness of the film was $L=256$~nm \cite{tag95}. This $d$ may be somewhat
underestimated since (i) the experimental slope is affected by pinning of
the domain wall, which we neglected, and (ii) Eq. (\ref{eq:dPdV}) somewhat
underestimates the exact result (Fig. 2). The result shows that already a
tiny passive layer has a profound effect on the hysteresis loop and its
slope rapidly diminishes with the growth of the passive layer. Thus, in thin
films with small lateral dimensions the transition into a fatigued state
with domains will be {\em discontinuous} and the growth of the passive layer
can easily trigger this transformation.

Within our model there is no hysteresis for an infinite film and the
dielectric response describes the slope of the hysteresis loop. It is
important to emphasize that the film becomes more `dielectrically rigid'
because of the growth of the passive layer alone, even without taking into
account either bulk or surface pinning of the domain walls. The estimated
slope of the fatigued system is in good correspondence with experimental
data, so that the growth of the passive layer might be the {\em main} {\em %
origin of fatigue.}

We believe also that the specific features of properties of the domain
structure in the presence of the passive layer may be a key to resolving the
famous {\em `paradox' of the coercive field}. All the estimates for creating
a nucleus with reversed polarization give {\em extremely large} values of
the energy barrier, $E^{\ast }\sim 10^{7}k_{B}T$ \cite{landau57,jan58,kay62}%
. Such a huge barrier makes a thermal nucleation impossible, and this makes
the related estimates of the coercive field
questionable\cite{landau57,jan58,kay62}. 

In our scenario, however, the coercive field is defined by the {\em energy
barrier} for the domain wall to {\em `enter'} the sample. It is
likely to be {\em much smaller }than the classic barrier for nucleation{\em %
\ }$E^{\ast }$, and this possibly resolves this long-standing paradox. We
hope that these new ideas about the problem of switching will prove
fruitful. For instance, we expect that in thin films the coercive field will
be increasing with decreasing {\em lateral} size of the sample, since the
size of a domain with a `wrong' polarization should become smaller than that
size. This should be contrasted with the usual conclusion that the coercive
field grows with decreasing {\em thickness} of the sample\cite{tag94},
irrespective of its lateral size. It would be very interesting to test this
prediction experimentally, especially in view of ever smaller sizes of films
used in devices.

The effect of the passive layer on switching in the ferroelectric films is a
specific case of a more general problem: the effect of inhomogeneities in
the system on its ferroelectric properties. As we have shown elsewhere \cite
{builtin}, a similar mechanism produces the fatigue of a film without any
actual passive layer, in presence of a depletion charge in
ferroelectric-semiconducting films\cite{Depl1}.

It is worth emphasizing that none of our conclusions depends on a
particular form for the domain structure (which may be of a checkerboard
type, etc.) This is because the effect of partial screening of the
depolarizing field by electrodes obviously applies to any domain pattern.

We appreciate useful discussions with J. Amano, G.A.D. Briggs, A. Gruverman,
and R.S. Williams. Extensive conversations with L. Wills-Mirkarimi were
invaluable. AL would like to acknowledge a support and hospitality of
Quantum Structures Research Initiative at Hewlett-Packard Labs (Palo Alto)
during the course of this work.


\begin{references}
\bibitem{land59}  M.E. Drougard and R. Landauer, J. Appl. Phys.{\bf \ 30},
1663 (1959).

\bibitem{dpin}  D.M. Smyth, Curr. Opinion in Sol. St. Mater. Sci. {\bf 1,}
692 (1996).

\bibitem{mil90}  S.L. Miller {\it et al.} J. Appl. Phys. {\bf 68}, 6463
(1990).

\bibitem{taylor95}  D.J. Taylor {\it et al.} Integr. Ferroelectrics {\bf 7},
123 (1995).

\bibitem{emig1}  H.M. Duiker {\it et al.} J. Appl. Phys. {\bf 68}, 5783
(1990); S.B. Desu and I.K. Yoo, Integr. Ferroelectrics {\bf 3}, \ 5783
(1993).

\bibitem{lem96}  V.V. Lemanov and V.K. Yarmarkin, Phys. Sol. State 38, 1363
(1996) [Fiz. Tverd. Tela {\bf 38}, 2482 (1996)].

\bibitem{tag96}  A.K. Kholkin {\it et al.} Appl. Phys. Lett. {\bf 68}, 2577
(1996).

\bibitem{gruv96}  A. Gruverman, O. Auciello, and H. Tokumoto, Appl. Phys.
Lett. {\bf 69}, 3191 (1996).

\bibitem{chensky82}  E.V. Chensky and V.V. Tarasenko, JETP {\bf 56}, 618
(1982) [Zh. Eksp. Teor. Fiz.{\bf \ 83}, 1089 (1982)].

\bibitem{LL8}  L.D. Landau and E.M. Lifshitz,{\it Electrodynamics of
Continuous Media }(Elsevier, 1985), (a)~Secs. 5, 10, 19.

\bibitem{LL44}  Ref.~\cite{LL8}, Sec. 44.

\bibitem{strukov}  B.A. Strukov and A.P. Levanyuk, {\it Ferroelectric
Phenomena in Crystals} (Springer, Berlin, 1998).

\bibitem{kit46}  C. Kittel, Phys. Rev. {\bf 70}, 965 (1946); T. Mitsui
and J. Furuichi, Phys. Rev. {\bf 90}, 193 (1953).

\bibitem{BLlong}  A.M. Bratkovsky and A.P. Levanyuk (to be published).

\bibitem{MagDomains}  Such a screening is absent in magnetic systems, hence
the behavior considered here has completely different origin compared to
magnetic domains [cf. A. Hubert and R. Sch\"{a}fer, {\it Magnetic Domains}
(Springer, Berlin, 1998)].

\bibitem{asymp}  $I$ has been transformed similar to \cite{LL44,kit46}
and $\sum_{n=1}^{\infty }b^{n}/n^{2}=\pi
^{2}/6-\int_{1}^{b}[\ln (1-x)/x]{\rm d}x$ was used to perform the series
expansion.

\bibitem{watan98}  Y. Watanabe, Phys. Rev. B{\bf \ 57}, 789 (1998); J. Appl.
Phys. {\bf 83}, 2179 (1998).

\bibitem{tag95}  A.K. Tagantsev {\it et al.} J. Appl. Phys. {\bf 78}, 2623
(1995).

\bibitem{landau57}  R. Landauer, J. Appl. Phys. {\bf 28}, 227 (1957).

\bibitem{jan58}  V. Janovec, Czech. J. Phys. {\bf 8}, 3 (1958).

\bibitem{kay62}  H.F. Kay and J.W. Dunn, Phil. Mag. {\bf 7}, 2027 (1962).

\bibitem{tag94}  A.K. Tagantsev {\it et al.}, Integr. Ferroelectrics {\bf 6}%
, 309 (1995); Integr. Ferroelectrics {\bf 4}, 1 (1994).

\bibitem{builtin}  A.M. Bratkovsky and A.P. Levanyuk, \prb {\bf 61},
Apr 1 (2000); cond-mat/ 9908070.

\bibitem{Depl1}  P.W. Blom {\it et al.} Phys. Rev. Lett. {\bf 73}, 2107
(1994).
\end{references}
\end{document}